\documentclass[%
 reprint,
 showkeys,
 showpacs,
 amsmath,amssymb,
 aps,
prb,
]{revtex4-1}

\usepackage{graphicx}
\usepackage{dcolumn}
\usepackage{bm}

\begin{document}

\preprint{APS/123-QED}

\title{Relation between the strength and dimensionality of defect-free carbon crystals\\}

\author{Sergey Kotrechko}
 \email{serkotr@gmail.com}
\affiliation{%
 G.V. Kurdyumov Institute for Metal Physics of the National Academy of Sciences of Ukraine, 03142 Kiev, Ukraine\\
}%
\author{Andrey Timoshevskii}
\affiliation{%
 G.V. Kurdyumov Institute for Metal Physics of the National Academy of Sciences of Ukraine, 03142 Kiev, Ukraine\\
}%
\author{Eugene Kolyvoshko}
\affiliation{%
 Taras Shevchenko Kyiv National University, Kyiv, Ukraine\\
 }%
\author{Yuriy Matviychuk}
\affiliation{%
 G.V. Kurdyumov Institute for Metal Physics of the National Academy of Sciences of Ukraine, 03142 Kiev, Ukraine\\
}%
\date{\today}

\begin{abstract}
On the basis of ab-initio simulations, the value of strength of interatomic bonds in one-, two- and three-dimensional carbon crystals is obtained. It is shown that decreasing in dimensionality of crystal gives rise to nearly linear increase in strength of atomic bonds. It is ascertained that growth of strength of the crystal with a decrease in it dimensionality is due to both a reduction in coordination number of atom and increase in the angle between the directions of atomic bonds. Based on these data, it is substantiated that the one-dimensional crystals have maximum strength, and strength of carbyne is the absolute upper limit of strength of materials
\end{abstract}


\maketitle


\section{\label{sec:level1}Background}
According to the existence paradigm of strength, there are two ways to increase strength, namely: (i) by increase in the density of defects in crystal lattice (increase in a lattice distortion), or, (ii) vice versa, creation of defect-free crystals. The first way is a typical approach to production ordinary ``bulk'' materials. High levels of the strength of nano-sized crystals are due, primarily, to the absence of defects ~\cite{1, 2, 3} . However, recently it was shown that extremely high levels of the strength and modulus of elasticity can be reached by decrease in dimensionality of crystals i.e., by the transition from three- to two- and one-dimensional crystals ~\cite{4, 5, 6}. Most clearly this effect can be demonstrated for allotropic forms of carbon. Thus, the strength of defect-free 3-D placeCityCrystal (diamond) is 95 GPa ~\cite{7}, of 2D-crystal (graphene) - 130 GPa ~\cite{8} and of 1D-crystal (carbyne) - more than 270 GPa ~\cite{9}. In this connection, the task of the work was to establish the regularities of influence of the crystal dimensionality on it strength and to develop ideas about the physical nature of this effect.

\section{\label{sec:level1}Methods}
One-, two- and three-dimensional carbon crystals were used as an object of investigation. To determine the strength of a one-dimensional crystal (monatomic chains of carbon atoms), \textit{ab-initio} calculations were employed. In this case, tension of the chain was simulated. Strength was determined for polyyne structure, which is energy-wise more favorable. Full energies of chains were calculated by the pseudo-potential method (software package \textit{Quantum-ESPRESSO} (QE)) ~\cite{10}. This method was used for modeling the mechanical properties of the chains. Pseudo-potentials for carbon were generated under the scheme \textit{Vanderbilt Ultrasoft} with the package \textit{Vanderbilt code}, version 7.3.4 ~\cite{11}. To determine the accuracy of the calculation of the total (full) energies of carbon chains, test calculations for infinite chains with the structure of polyynes were executed. Interatomic distances and total  energies, consistent with the results of the work ~\cite{12} were obtained. The value of the cutoff energy \textit{Ecut = 450 eV}. In the calculations, the exchange-correlation potential PBE ~\cite{13} was used.

To determine the strength of atomic bonds, $R_{2D}^{} $, in two-dimensional (2D) crystal (graphene), the value of fracture stress of graphene ~\cite{14} obtained by \textit{ab-initio} simulation of tension of graphene sheets in two directions ``zigzag'' and ``armchair'' were used. The magnitude of atomic bond strength \textit{Fc} was determined as the value of critical force acting on the bond at the time of instability of the graphene sheet at it tension:

\begin{equation} \label{EQ__1_} R_{2D}^{zig} =\sigma _{c}^{zig} d_{0} b_{o} \times \left(1-\varepsilon _{} \right) \end{equation}

\begin{equation} \label{EQ__2_} R_{2D}^{arm} =\sigma _{c}^{arm} d_{0} b_{0} \times \left(1-\varepsilon _{} \right)\times \left(1+cos\left(\theta _{0} \right)\right) \times cos\left(1-\theta'\right) \end{equation}

where $\sigma _{c}^{zig} $ and $\sigma _{c}^{arm} $ are the critical stress of instability of graphene sheet under tension in the directions "zigzag" and "armchair", respectively; $b_{{\rm 0}} =\sqrt{3} /2a_{{\rm 0}} $; $a_{0}=0.1422\: nm$ is the lattice parameter; $d_0=0334\: nm$ is the effective thickness of graphene sheet; $\theta _{0}=120^\circ$ is the angle between atomic bonds $ij$ and $jk$ in undeformed state; $\theta'=130^\circ$ is the angle, at which instability in graphene occurs (was determined based on the geometry of the system at instability); $\varepsilon $ is the transverse deformation of graphene sheet.

The strength of atomic bonds in three-dimensional (3D) crystal (diamond) was determined on the basis of \textit{ab-initio} simulations of tension of diamond crystals in the direction $<111>$ ~\cite{7}. The strength value was calculated as the critical magnitude of the force acting on the interatomic bond at instability of the crystal under tension:

\begin{equation} \label{EQ__3_} R_{3D} =\frac{3\sqrt{3} }{2} \sigma _{c} a_{{0}}^{{2}} {cos}^{{2}} \left(\pi /{2}-\theta \right) \end{equation}

where $\sigma _{c}$ is the critical stress of instability , $a_0$ is the lattice parameter, $\theta $ is the angle between atomic bonds $ij$ and $jk$ in deformed state.

\begin{figure}
\includegraphics{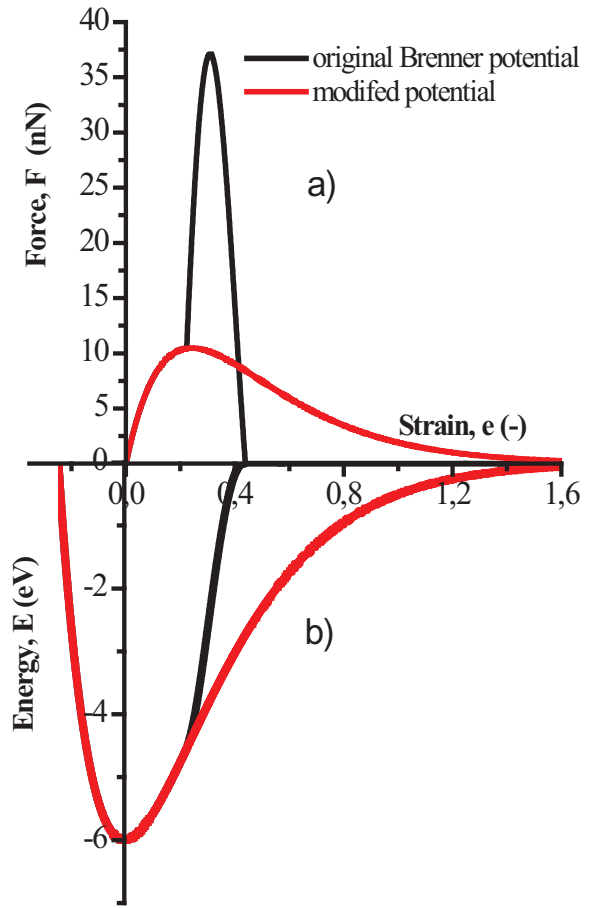}
\caption{\label{fig:fig1} (Color online) The original and modified Brenner potential. Force vs. strain (à). Energy vs. strain (b).
}
\end{figure}

Theoretical analysis of the orientation dependence of strength and its value dependence on coordination number of atom was executed within the formalism of Brenner potential ~\cite{15}, which is based on Abel pseudo-potential theory ~\cite{16} that describes well the carbon compounds. Parameterization of the potential was carried out by the set II of Brenner parameters ~\cite{15}: $R=0.139\: nm$, $D(e)=6.0$, $s=1.22$, $\beta =0.21\: nm^{-1}$, $\delta =0.5$, $a_0=0.00020813$, $c_0^2=330^2$, $d_0^2=3.5^2$. Since the Brenner potential in initial state gives a significant overestimation of strength of atomic bonds, which reaches values of 30nN (Fig. ~\ref{fig:fig1}a) , its modification was carried out for the crystal that consisted in correction of cutoff function $f_{cut} (r_{ij}) \equiv 1$. It enabled to get the values of strength, which agree well with the results of \textit{ab-initio} simulations.

\section{Results and discussion}
\begin{figure}
\includegraphics{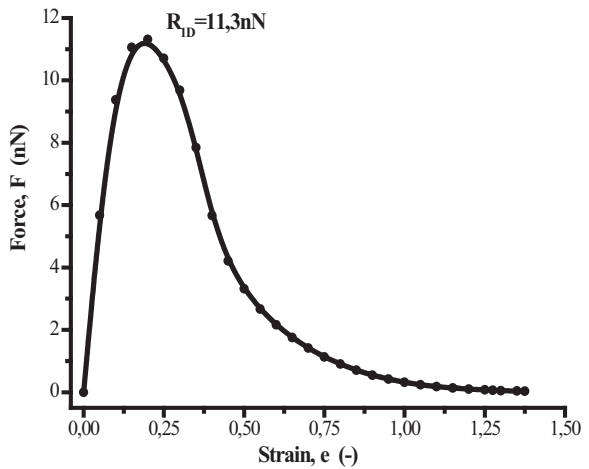}
\caption{\label{fig:fig2} (Color online) Dependence "force vs. strain" for infinite monatomic chain. Result of \emph{ab-initio} calculations.
}
\end{figure}

The strength of atomic bonds in one-dimensional crystal was determined by the critical stress of instability of monatomic chain under the conditions of uniaxial tension (Fig.~\ref{fig:fig2}). In accordance with the data obtained, $R_{1D} =11.3\: nN$. This value is in good agreement with the evidence of 12.2 nN ~\cite{17}, and with the value 12.3 nN, obtained for chains containing more than ten atoms ~\cite{4}.

\begin{figure}
\includegraphics{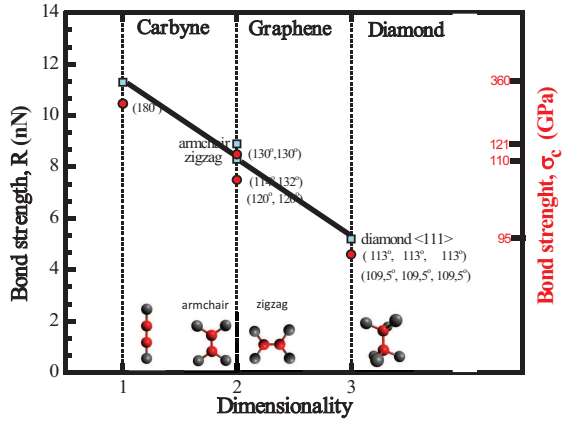}
\caption{\label{fig:fig3} (Color online) Dependence of the strength of interatomic bonds of carbon crystals on the its dimensionality: $\blacksquare$ are the strength values obtained by the results of \emph{ab-initio} calculations; and $\bullet$ are the results of calculation by the modified Brenner potential for angles between interatomic bonds in the unstrained state and at the critical strain of crystal instability, respectively.
}
\end{figure}

To determine the strength of atomic bonds in 2D-crystals, data of \textit{ab-initio} simulation of tension and fracture of graphene sheets in "zigzag" and "armchair" direction ~\cite{14} were used. The critical value of the force at the moment of instability of the graphene sheet was determined by equations \eqref{EQ__1_} and \eqref{EQ__2_}. According to the results obtained, the strength of atomic bonds in graphene for "armchair"  direction was 8.9 nN and for "zigzag" direction - 8.3 nN. It should be noted that the difference in angles between atomic bonds in the stretched sheet of graphene is the cause for different values of atomic bonds strength in graphene at transition from "zigzag" to "armchair" orientation. Thus, at the moment of instability under tension in "armchair" direction the orientation of atomic bonds is characterized by angles of $130^\circ$ and $130^\circ$, while for tension in "zigzag" direction, they are $114^\circ$, $132^\circ$, respectively (Fig. ~\ref{fig:fig3}).  Really, calculations using Brenner potential, in the first case give the value of strength of atomic bonds equal to 8.49 nN, and in the second case this value is 7.5 nN (Fig.~\ref{fig:fig3}). Accordingly, the ratio of the strengths of atomic bonds obtained according to \textit{ab-initio} calculations is 1.07, and of those obtained from the Brenner potential is 1.14. The difference does not exceed 9\%. The absolute values of these strength "anisotropy" are not much great, but they illustrate one of the specific features of interatomic interaction of carbon atoms, namely, its orientation dependence.

The value of bond strength in diamond was obtained based on the results of \textit{ab-initio} simulations. The critical value of the force acting in the bonding at the time of instability of the diamond lattice was determined by the formula \eqref{EQ__3_}. In compliance with the results obtained bond strength in diamond is $R_{3D}=5.2\: nN$.

According to the results, change in the dimensionality of defect-free crystals essentially effect on their bond strength. This effect manifests itself in a significant (more than 2 times) increase in the bond strength of the transition from three- to two- and one-dimensional crystals (Fig. ~\ref{fig:fig3}). Strength values of bonds calculated using the modified Brenner potential are plotted on the same graph. Calculations were executed for both the initial crystal and taking into account changes of angles $\theta $ between the bonds when reaching a critical strain of crystal instability. The results of calculations, agree well with the \textit{ab-inito} results. As it is indicated in Fig.~\ref{fig:fig1}, the transition from three-dimensional to two- and one-dimensional crystals, change in coordination numbers is accompanied by a change in the angles $\theta $ between the directions of atomic bonds. To separate the contributions of the coordination number Z and the angle $\theta $ to this change in the strength, $R$, Brenner formalism was used. In explicit form the dependence of strength $R$ on the coordination number $z$ may be derived only for the case of equal angles $\theta _{ijk} $ with all neighboring atoms $k$:

\begin{equation} \label{EQ__4_} R(Z,\theta)=D^{(e)} \beta \frac{\sqrt{2s}}{s-1} \left(s^{\frac{1}{1-s}}-s^{\frac{s}{1-s}}\right)\left(1+(z-1)G\left(\theta_{ijk} \right)\right)^{\frac{\delta s}{1-s} }  \end{equation}

where $G_i(\theta_{ijk})$ is angle function ~\cite{15}.

\begin{figure}
\includegraphics{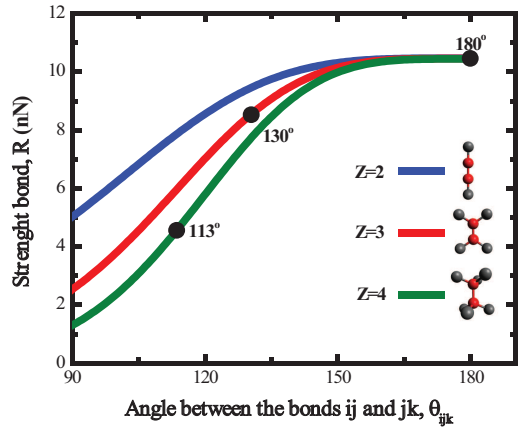}
\caption{\label{fig:fig4} (Color online) Influence of bonds orientation on the strength of carbon crystals at a fixed value of coordination number Z: $\bullet$ – angles between the bonds at the moment of instability of carbyne ($180^\circ$), graphene ($130^\circ$) and diamond ($113^\circ$).
}
\end{figure}

Using the relations suggested in ~\cite{15} and the magnitudes of  Brenner parameters (set II from ~\cite{15}), orientation dependencies of strength at fixed values of coordination number $Z=2, 3, 4$ were built (Fig.~\ref{fig:fig4}).

Formally, Brenner dependence describes the change in strength with the angle within the range from $0^\circ$ to $180^\circ$. But in the real objects (diamond, graphene, carbyne) variation of the angle does not exceed the range $90^\circ - 180^\circ$ deg, so the analysis is confined to this range of values.

\begin{figure}
\includegraphics{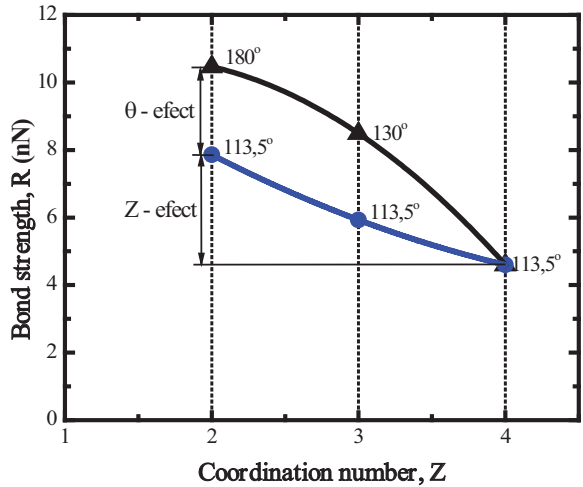}
\caption{\label{fig:fig5} (Color online) Dependence of the atomic bond strength on the coordination number at the fixed values of angles. $\blacktriangle$ are the critical value of angles.
}
\end{figure}

According to the data shown in Fig. ~\ref{fig:fig4}, the sensitivity of strength to change in the angle increases with the growth of coordination number \textit{Z}. On the other hand, the sensitivity of strength to changes in coordination number increases with decreasing in angle. As it is indicated in Fig. ~\ref{fig:fig5}, at transition from diamond to carbyne, bond strength increases $2,3$ times, while 56\% of this increase in the strength is due to a decrease of the coordination number and 44\% is related to the growth of angle to the maximum possible one, which is $180^\circ$.

Regularities of dependence of atomic bond strength on coordination number is demonstrated by the carbon, because it is most convenient to analyze this problem. However, these regularities are inherent also to the other materials. Work ~\cite{18} gives evaluation of the strength of atomic bonds in 1D and 3D crystals for metals such as Cu, Ag, Au, Pd and Pt. The results of calculations by different methods shown that at transition from 1D- to 3D-crystals, the strength of atomic bonds should increase by $2 - 3.3$ times. Thus, the material in the form of a one-dimensional crystal (monatomic chain) should have the highest level of strength. In this regard, strength of carbyne claims to be the absolute upper limit of strength of materials. According to the results of \textit{ab-initio} calculations, carbyne containing 5 atoms has a maximum strength. This strength is 13.1 nN ~\cite{4}. When the effective diameter of the chain equals $0.200\: nm$ ~\cite{9}, its value is 417 GPa. In ~\cite{19} for an infinite chain the strength values $9.3-11.7\: nN$ were obtained, and considerably smaller value of effective diameter $d = 0.077\: nm$ was utilized. This gives the strength value for infinite chain equal to $1988-2501$ GPa. In ~\cite{9} the experimental measurement of carbyne strength was executed. It was ascertained that its strength must exceed 270 GPa. This lower bound for strength of carbyne is more than 2 times greater than the experimental value of strength of graphene (two-dimensional crystal of carbon).

\section{Conclusions:}
Increase in strength of the interatomic bonds while reducing in dimension of the crystal is a specific feature of nano-sized defect-free crystals. At transition from three to two and one-dimensional crystal of carbon, interatomic bond strength increases almost linearly. The bonding strength in carbon monatomic chain is 2.2 times higher than that in diamond. This value is of the same order as the increase in strength of metallic crystals, which may be 2-3 times.

Growth of crystal strength with decreasing in its dimensionality is due to two factors, namely, a decrease in coordination number (the number of atomic bonds) and increase in the angle between the bonds. At transition from diamond to carbyne, 56\% of increase in strength is due to a decrease in coordination number and 44\% - due to increasing in angle between the bonds.

\begin{acknowledgments}
The authors gratefully acknowledge the financial support from Project \# 15/14-Í. Program "NANOSYSTEMS, NANOMATERIALS AND NANOTECHNOLOGIES" of NAS of the Ukraine.
\end{acknowledgments}

\appendix
\section{Competing interests}
The authors declare that they have no competing interests.

\section{Authors' contributions}
\textbf{SK} formulated the research problem and managed investigations. \textbf{EK} executed the calculation of the critical value of interatomic interaction force for 3D-crystal (diamond) and 2D-crystal (graphene); carried out a theoretical analysis of the orientation dependence of strength and dependence of coordination number. \textbf{AT} and \textbf{YuM} carried out \textit{ab-initio} calculations. All authors analyzed the results and participated in the preparation of the manuscript. All authors read and approved the final manuscript.

\section{Authors' information}
\textbf{SK} is Doctor of Sciences in Physics and Mathematics, Professor, Head of Department of Strength and Fracture Physics, highly qualified specialist in multi-scale approach to strength and reliability of materials.

\textbf{AT} is Ph.D. in Physics and Mathematics, Senior Researcher, eminently qualified expert in computational solid state physics.

\textbf{EK} is BSc in Physics, graduate of the Faculty of Physics of National Taras Shevchenko University of Kyiv, researcher of the Department of Physics of Metals.

\textbf{YuM} is Ph.D. in Physics and Mathematics, Senior Researcher, specialist in computational solid state physics.


\bibliography{chains_pap3}

\end{document}